\documentclass{sig-alternate-per}

\usepackage{algorithm}
\usepackage[noend]{algorithmic}

\usepackage{amsmath}
\usepackage{amssymb}
\usepackage{graphicx}
\usepackage{hyperref}
\usepackage{algorithm}
\usepackage{algorithmic}
\usepackage{comment}

\newcommand{\ALG}{\textsc{Tetris}}
\newcommand{\OKTO}{\textsc{Oktopus}}

\newcommand{\pricing}{\textsc{DSP}}

\title{Opposites Attract: \\ Virtual Cluster Embedding for Profit}

\author {
   Arne Ludwig$^1$,
   Carlo Fuerst$^1$,
   Alexander Henze$^1$,
   Stefan Schmid$^{1,2}$\\
\small{$^1$ TU Berlin, Germany, $^2$ Telekom Innovation Laboratories (T-Labs),
Germany}\\
\small{\texttt{\{arne,carlo,ahenze\}@inet.tu-berlin.de, stefan@net.t-labs.tu-berlin.de}}\\
}

\begin{document}
\maketitle

\sloppy

\begin{abstract}
It is well-known that cloud application performance critically depends
on the network. Accordingly, new abstractions for cloud applications are
proposed which extend the performance isolation guarantees to the network.
The most common abstraction is the \emph{Virtual Cluster} $VC(n,b)$: the $n$ virtual
machines of a customer are connected to a virtual switch at bandwidth $b$.
However, today, not much is known about how to efficiently embed and price
virtual clusters.

This paper makes two contributions. (1) We present an algorithm called $\ALG$ that
efficiently
embeds virtual clusters arriving in an online fashion, by jointly optimizing the node and link resources.
We show that this algorithm allows to multiplex more virtual clusters over the same physical
infrastructure compared to existing algorithms,
hence improving the provider profit. (2)
We present the first demand-specific pricing model called $\pricing$ for virtual clusters. Our pricing model is fair in the sense
that a customer only needs to pay for what he or she asked. Moreover, it
features other desirable properties, such as price independence from mapping locations.
\end{abstract}

\section{Introduction}

Server virtualization has revamped the server business over the last years,
and cloud computing has changed the way we think about resource allocation in the Internet.
However, while cloud providers support the flexible allocation of virtual machines (\emph{VMs}) and
provide the customer with certain resource and performance isolation guarantees,
networking has so far been treated as a second class citizen: until recently,
cloud customers could not specify even basic networking guarantees to connect their virtual
machines.

This is problematic given the fact that the traffic generated by cloud applications such as MapReduce and distributed
databases
is not negligible. Indeed, it has been shown that cloud applications suffer from resource interference
on the network, in the sense that application performance can become unpredictable. Longer job execution times
also entail higher costs for the customers who are charged on a \emph{per-VM-hour basis}.~\cite{short-talk-about}

To overcome these problems, more powerful resource reservation models have been proposed.~\cite{short-ballani2011towards}
A common abstraction is the \emph{Virtual Cluster}: a virtual cluster $VC(n,b)$ connects
$n$ virtual machines to a virtual switch at bandwidth $b$---essentially a hose model.~\cite{short-talk-about}
While the resulting performance isolation guarantees are attractive, the virtual cluster abstraction
raises two fundamental questions: (1) How to embed virtual clusters on a given physical network? In order to
make an optimal use of its resources and hence maximize the profit, a provider may want to multiplex as
many virtual clusters as possible over the physical infrastructure (while fulfilling the requested virtual
cluster specification). (2) How to efficiently price virtual clusters?
While today's cloud pricing models typically focus on VM hours only, the pricing of virtual clusters becomes
a 2-dimensional problem: a customer requesting more bandwidth should be priced accordingly.
This paper addresses these two questions.

\textbf{Our Contribution.} We present the first pricing scheme for virtual clusters, called $\pricing$ (\emph{Demand-Specific Pricing}),
which explicitly takes into account the different computation and bandwidth requirements. That is, unlike
the dominant-resource pricing scheme $\emph{DRP}$ presented in the literature before~\cite{short-ballani2011price},
$\pricing$ is designed according to a specification-dependent, \emph{pay-only-for-what-you-request policy}:
While in $\emph{DRP}$, the size $n$ and the bandwidth $b$ of a virtual cluster are strictly coupled,
$\pricing$ allows customers to request virtual clusters $VC(n,b)$ of
arbitrary and independent size $n$ and bandwidth $b$, and be priced accordingly and in a fair manner.
Moreover, $\pricing$ ensures desirable properties such as location independent pricing.

Together with this pricing scheme we present a new embedding algorithm called $\ALG$\footnote{The
name of the algorithm is due to its strategy to balance different resources equally, see also Figure~\ref{fig:imbalance-pic}
for an illustration.} which
is also specification-dependent in the sense that $\ALG$ accounts for differences in the node and link requirements of virtual clusters.
Concretely, $\ALG$ is tailored to an online scenario where different virtual clusters $VC_1(n_1,b_1),VC_2(n_2,b_2),\ldots$ are
requested over time, and collocates ``opposites'': computation-intensive but
communication-extensive virtual clusters are mapped together with
computation-extensive but communication-intensive virtual clusters, to maximize the number of simultaneously hosted
virtual clusters over time.
Given the online nature of the problem, this is a non-trivial task.
We show that our algorithm outperforms previous algorithms, in the sense that
a provider can host more virtual clusters, and hence increase its profit.

\section{Background \& Model}\label{sec:model}

Most cloud providers today still offer virtual machine services only,
charging their customers on a per-hour basis.
However, we witness a trend towards more network oriented specifications
(see also, e.g., \emph{Amazon Placement Groups}, \emph{Amazon EBS-Optimized Instances} or \emph{Microsoft Azure ExpressRoute}),
and especially the virtual cluster abstraction is becoming a
popular model for datacenter applications.~\cite{short-ballani2011towards}

In~\cite{short-ballani2011towards}, a first algorithm (henceforth called $\OKTO$) was proposed
to embed virtual clusters in fat-tree datacenter topologies,
and Ballani et al.~\cite{short-ballani2011price} proposed a first pricing scheme
for virtual clusters which give minimal bandwidth guarantees.
Essentially, their scheme is based on \emph{Dominant Resource Pricing}, short \emph{DRP}.
\emph{DRP} provides different templates for the customers, with predefined sizes $n$ and
an associated amount of minimal guaranteed bandwidth $b$.
While this model is attractive for its simplicity,
also in the sense that the interface between the customer and provider may not have to be changed,
the minimal bandwidth $b$ is a function of $n$ and cannot be chosen by the customer.
As such, \emph{DRP} is still 1-dimensional and does not leverage the full flexibility of the virtual cluster model,
which is described by two
\emph{independent} parameters $n$ and $b$.

Indeed, virtual cluster specifications are likely to come with different requirements~\cite{short-mesos}
and can be heterogeneous~\cite{short-talk-about}: a latency sensitive webservice
can be very different in nature than, say, a delay-tolerant batch processing job or a network-hungry database synchronization application.
One implication of the \emph{DRP} scheme is that customers who know their virtual cluster demands might suffer
from the inherent 1-dimensionality: in order to meet their application requirements,
customers may be forced to upgrade to the next larger template, increasing both resources.

This paper seeks to overcome this problem by allowing customers to specify their computation
and communication requirements separately.
Concretely, we allow the customer to specify three parameters independently:
the number of VMs $n$, the computational type $c$ of the virtual machines (e.g., small, medium, or large instances),
as well as the bandwidth $b$. That is, we use a virtual cluster abstraction $VC(n,c,b)$,
where all virtual machines are of the same computational type $c$, and are connected to a virtual switch at bandwidth $b$.

We use the following conventions in our notation. The computational type $c$ is normalized in the sense that $c$
describes the fraction of the capacity of a physical server. Similarly, we will normalize
$b$ to denote the requested fraction of the overall link capacity of a physical server.
A central concept for our algorithm $\ALG$ (improving upon $\OKTO$) and pricing scheme $\pricing$ (improving upon \emph{DRP}) is the \emph{resource ratio} between
the requested node and link resources, henceforth denoted by $\rho = c/b$.

In general, we will assume that requests arriving in an online fashion have to be immediately embedded or rejected by the
provider. In order to successfully embed a virtual cluster, the provider has to fulfill all its specifications.

\section{Pricing Scheme}\label{sec:pricing}

The proposed specification-dependent pricing scheme $\pricing$ is based
on a unit price for computation, henceforth denoted by $p_c$, as well as a unit price
for communication (i.e., bandwidth), henceforth denoted by $p_b$.
Ideally, for a virtual cluster request with a per-VM computational demand $c$
and a per-VM bandwidth demand $b$, a customer should be charged according to the resource
proportions, e.g.
$$P_{\emph{ideal}}=n\cdot (c \cdot p_c + b \cdot p_b)$$

However, compared to a dominant resource pricing scheme,
this solution can result in a lower income at the provider,
especially if requests are highly heterogeneous leading to a higher fragmentation.
While this income loss could be compensated by increasing
the unit prices $p_c, p_b$
accordingly, one has to be careful not to punish customers with an ideal resource ratio $\rho=1$,
who would prefer providers offering \emph{DRP} in this case.
To solve this problem, in the following, we propose to add a small charge for customers with a resource ratio $\rho \neq 1$.

But let us first revisit the \emph{DRP} scheme given our notation.
In \emph{DRP}, a customer who requests a virtual cluster $VC(n,c,b)$, with relatively lower resp.~higher computation
requirements compared to the communication requirement, is forced to upgrade the request to the next larger
template for both resources.
The corresponding formula for the \emph{DRP} scheme is
$$P_{\emph{DRP}}=n\cdot \left[\max\{c, b\}\cdot (p_c + p_b)\right]$$

In order to bridge the difference between $P_{\emph{ideal}}$ and $P_{\emph{DRP}}$,
we propose the following
\emph{demand specific pricing} scheme
$\pricing$ which introduces an extra fee for requests with an unbalanced resource ratio $\rho$.
In this paper, we will assume a linear dependency between the extra cost and $\rho$, although other dependencies (e.g.,
quadratic) could
also be expressed in our model. This decision is based on the assumption that more skewed requests are more likely to
generate fragmentation. In summary, $\pricing$ computes the virtual cluster price as follows:
 $$P_{\emph{DSP}}= n\cdot (b \cdot p_b + c \cdot p_c) + \left\{\begin{array}{cl}
(c-b)\cdot p_b \cdot \lambda_b, & \mbox{if }c\geq b\\ (b-c)\cdot p_c \cdot \lambda_c, & \mbox{else} \end{array}\right.$$
\noindent where $\lambda_c, \lambda_b\geq 0$ are weighing factors.
Note that this scheme can be seen as a generalization of the dominant resource pricing strategy:
$\lambda_c,\lambda_b=1$ implies that $P_{\emph{DSP}}=P_{\emph{DRP}}$.
Lower weights result in savings for the customers, and $\lambda_c,\lambda_b=0$ implies
$P_{\emph{DSP}}=P_{\emph{ideal}}$.
Given that
the customers have a good understanding of their specific requirements in terms of computation and
communication---a reasonable assumption as we argue---this pricing scheme leads to a higher provider profit, and in a
competitive market,
the extra income compared to \emph{DRP}
is shared with the customers.

Let us elaborate on the weighing factors $\lambda_c$ and $\lambda_b$.
In general, the factors should depend on the amount of embedded virtual requests as well as the current resource demand
and supply.
If the provider has a good estimation of the virtual clusters that will be requested,
the values can be computed ahead of time; otherwise, the factors may be estimated over time, see~also
Section~\ref{sec:simulations}
for a discussion.
Given a difference of $\Delta$ between the provider profit under \emph{DRP} for upgraded and non-upgraded
requests, the extra income generated by the lower resource consumption of the non-upgraded requests could be evenly
distributed over requests requiring more of either one of the two resources.
(Recall that the price of requests with $\rho=1$ will not change.)
The calculation for $\lambda_b$ in a scenario with $N$ virtual machines for which $c>b$ and with
expected
requirements $E(c)$ and $E(b)$, is given by
$$N \cdot (E[c]-E[b]) \cdot p_b \cdot (1-\lambda_b) = \Delta/2$$
\noindent The factor $\lambda_c$ can be computed similarly.
In both cases, $c>b$ and $b>c$, a similar difference for $E[c]$ and $E[b]$ also leads
to similar fees. This is fair, as one type of request only generates more
profit because of the other one.

Also note that \pricing~keeps the desirable location independence of \emph{DRP}~\cite{short-ballani2011price}.
However, unlike \emph{DRP}, \pricing~is specification-dependent, i.e., a customer only has to pay
for what he or she specified.

\section{Embedding Algorithm}\label{sec:algo}

In order to fully exploit differences in the virtual cluster specification and in order
to maximize the resource utilization (and hence provider profit),
a new embedding algorithm has to be devised: the state-of-the-art virtual cluster
embedding algorithm, Oktopus~\cite{short-ballani2011towards} (as well as its variants~\cite{short-proteus}), are
based on an aggressive collocation strategy, which turns out to be suboptimal in settings where requests can have
resource ratios $\rho\neq1$.

\begin{figure}[t]
  \centering
\includegraphics[width=.95\columnwidth]{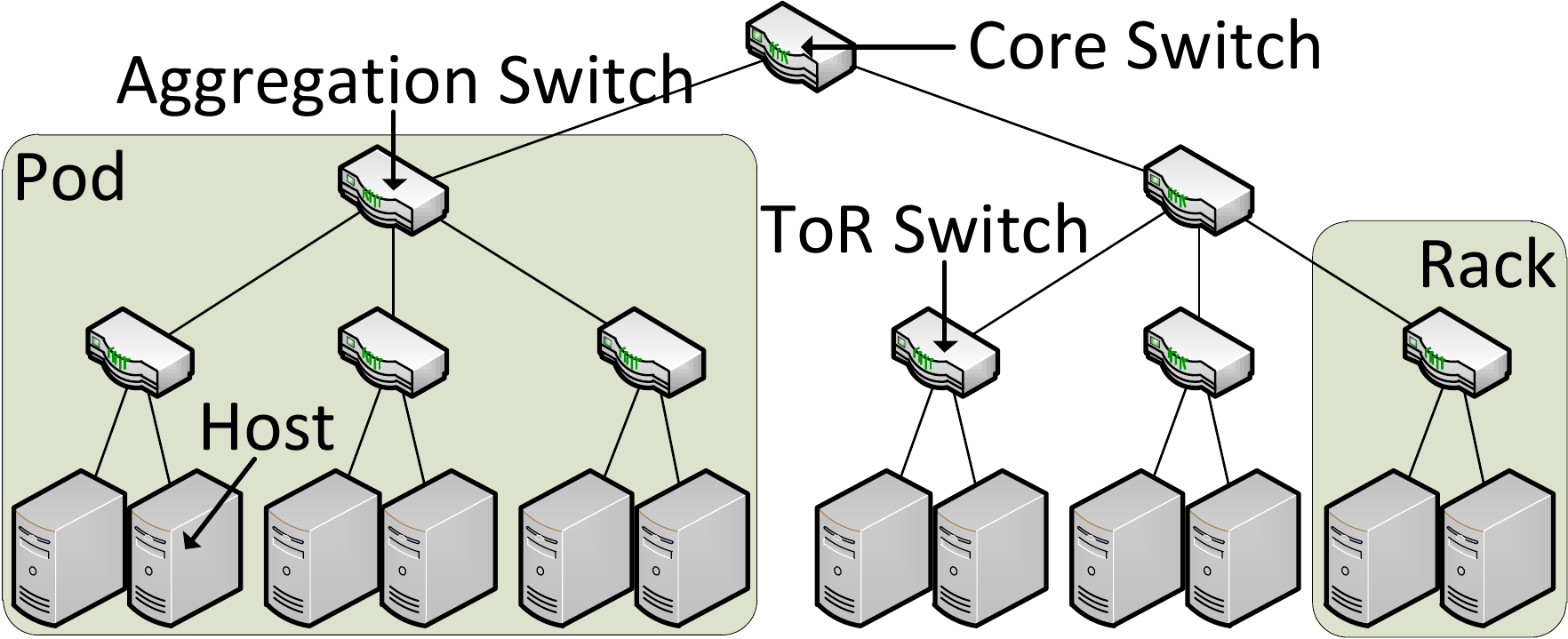}
      \caption{Fat-tree datacenter topology.}
      \label{fig:fattree-pic}
  \end{figure}

In the following, we propose $\ALG$, a virtual cluster embedding algorithm which leverages the virtual cluster
specification details.
 The algorithm is tailored toward fat-tree datacenter topologies (cf~Figure~\ref{fig:fattree-pic}),
 the standard architecture today.
In a nutshell, hosts (or equivalently: servers) which are connected to the same top of rack (ToR) switch, constitute a
rack.
Racks connected to the same aggregation switch form a pod. The fat-tree depicted in Figure~\ref{fig:fattree-pic}
consists of two pods,
containing three racks each; there are two hosts per rack.

\begin{figure}[ht]
  \centering
\includegraphics[width=.95\columnwidth]{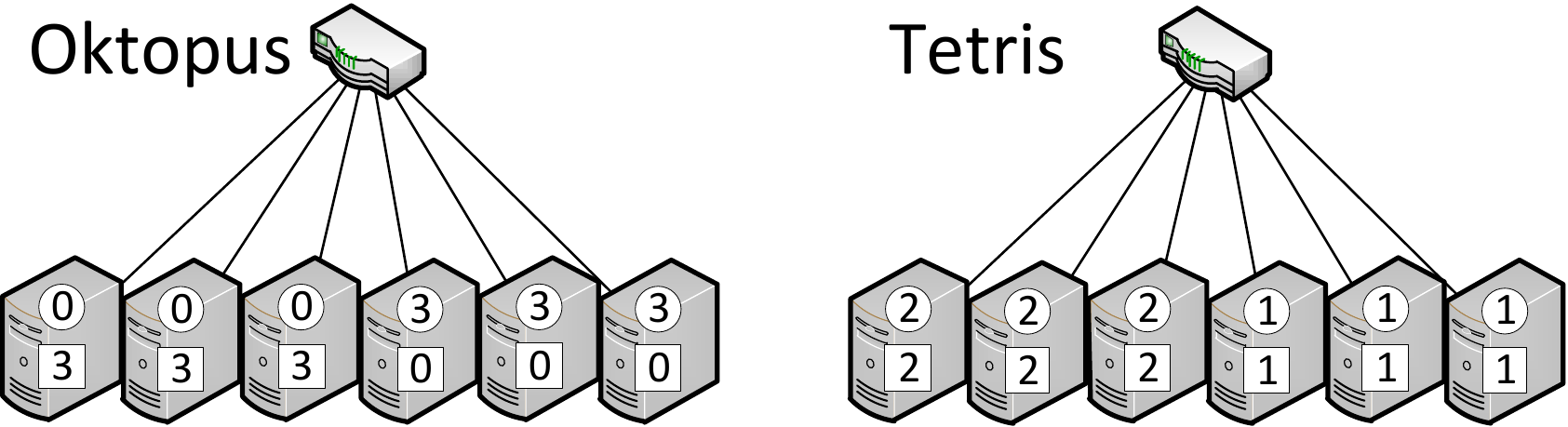}
      \caption{Embedding behavior of $\OKTO$ and $\ALG$. Six hosts are connected to a switch. Two $VC$s are requested: $VC_1(9,1/6,2/6)$ and $VC_2(9,2/6,1/6)$. The numbers in the
circles represent \emph{VMs} of $VC_1$ which are mapped to the corresponding hosts, the numbers in the squares represent
\emph{VMs} of $VC_2$.}
      \label{fig:imbalance-pic}
  \end{figure}

We will compare our embedding algorithm $\ALG$ to the $\OKTO$ algorithm~\cite{short-ballani2011towards}. $\OKTO$ is designed to embed arbitrary virtual clusters efficiently and is
not limited
to any templates. Hence it can also directly be used as an embedding algorithm for \pricing, without any modifications.
To find an embedding, $\OKTO$ traverses the different hierarchical levels of the fat-tree. It tries to
embed
the virtual cluster on single hosts first, and if no solution is found, it continues on the rack level. This process is
repeated until a
solution is found or $\OKTO$ failed to find a solution over the entire substrate. As a result of this approach, the
resulting embeddings are dense and use low amount of bandwidth.
The problem of such dense embeddings is that requests with a resource ratio $\rho \neq1$ are collocated which wastes
physical resources. Figure~\ref{fig:imbalance-pic} (\emph{left}) illustrates this point.
For \OKTO, $VC_1$ is embedded on the right three hosts. The residual capacity in terms of VM slots on each host is $1/2$ of its total capacity,
however, the bandwidth on the links is used up, rendering it unlikely that the remaining VM slots
can be used in the future. On the other hand, the left three hosts, which host $VC_2$, only utilize $50\%$ of their bandwidth, but have
no remaining VM
slots.

The core idea of our algorithm $\ALG$ is to distribute
skewed $VC$s over multiple hosts, without increasing the bandwidth costs compared to \OKTO. Similar to \OKTO, $\ALG$
also
traverses the hierarchical levels (short: $\ell$) of the fat-tree, but instead of collocating as many \emph{VMs} on a
single host as
possible, $\ALG$ distributes the \emph{VMs} over physical machines (short: $p$) depending on the ratio of the residual
resources per host. This is described
in Algorithm~\ref{alg:rob}.

\begin{algorithm}[ht]
    \caption{\ALG}
    \label{alg:rob}
    \begin{algorithmic}[1]
    \REQUIRE Fat-tree $T$, virtual cluster $VC$
    \FOR{$\ell \in \{$host($T$), rack($T$), pod($T$), root($T$)$\}$}
      \FOR{$v \in VC$}
	\STATE find $p \in$ $\ell$ with highest ratio of residual resources after embedding of $v$
      \ENDFOR
      \IF{$\forall~ v$ embedding found}
	\RETURN embedding	
      \ENDIF
    \ENDFOR
    \RETURN $\bot$	
    \end{algorithmic}
\end{algorithm}

In our example in Figure~\ref{fig:imbalance-pic}, using $\ALG$
results in a distributed embedding of both
$VC$s. While all resources on the left three hosts are utilized, the right three hosts have spare capacities, both in
terms of bandwidth and \emph{VM} slots. Hence subsequent requests can more likely be accepted.

Note that the current design of $\ALG$ only considers the bandwidth on the access
level. Hence, it can fail to find a feasible solution if the other layers of the fat-tree are oversubscribed. Therefore
our current implementation treats $\ALG$ as an extension to $\OKTO$ and falls back to regular $\OKTO$ behavior if no
solution was found.

\section{Simulations}\label{sec:simulations}

In order to study the benefits and limitations of $\pricing$ and $\ALG$ in different settings,
we implemented a discrete event simulator. As the pricing results also
depend on the embedding algorithm, we study three combinations:  we integrated \emph{DRP} with $\OKTO$, $\pricing$ with
$\OKTO$, and
\emph{\pricing} with $\ALG$. To ensure a fair comparison, we use the same parameters and methodology
as in~\cite{short-ballani2011towards} and~\cite{short-ballani2011price}.

\textbf{Requests.} The virtual cluster requests arrive according to a Poisson process with exponentially distributed
durations, chosen to induce a system load of around $80\%$. By default, the mean size of a $VC$ is $n = 49$, $c$ and $b$
are
chosen uniformly from $\{1/8,1/4,1/2\}$. The templates $(c,b)$ of \emph{DRP} are $(1/8,1/8), (1/4,1/4), (1/2,1/2)$ and the customer is
bound to select the next larger template for requests with $\rho\neq1$. For each parameter
set, we request $80k$ virtual clusters. To avoid artifacts related to the
initially empty datacenter, we start evaluating our metrics after $10k$ requests. The remaining values are omitted from
the dataset.

\textbf{Physical Setup.} We model our datacenter as a three-layer fat-tree (Figure \ref{fig:fattree-pic} illustrates a
small fat-tree). $40$ hosts form a rack, $40$ racks form a pod. In total we have $10$ pods. Given that each physical
element has a capacity of $8$  \emph{VM} units, this leads to a total capacity of $128k$ small VMs. By default we assume
that the links between the ToR switches and the aggregation switches are oversubscribed by a factor of $4$, while the
links between the aggregation switches and the core are not oversubscribed.

\textbf{Metrics.} Various works have used the acceptance ratio of an embedding algorithm in order to measure its
performance. This metric however, is biased to prefer algorithms which accept a large number of small requests instead
of fewer bigger ones. Therefore, we decided to evaluate the sum of the embedded virtual resources in both dimensions:
bandwidth and VM slots. A request $VC(10,1/4,1/8)$ will have a \emph{resource sum} of $10$ for bandwidth and $20$
for \emph{VM} slots. Note that even though we will embed a $VC(10,1/4,1/4)$ for \emph{DRP}, the bandwidth sum remains $10$,
as the customer does not benefit from the over-provisioning.

\begin{figure}[ht]
  \centering
\includegraphics[width=.95\columnwidth]{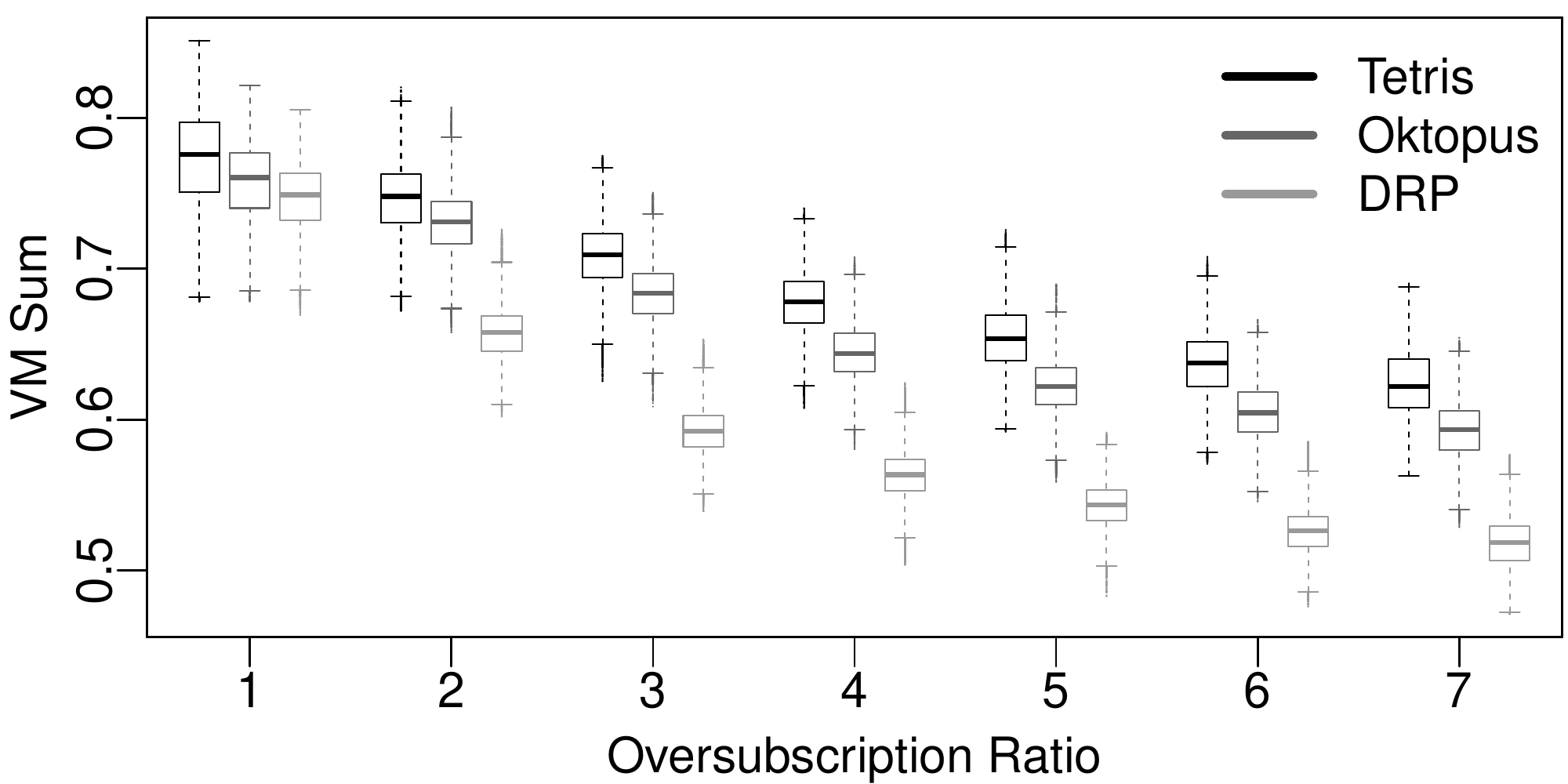}
      \caption{Embedded \emph{VM} slots for $\ALG$ with $\pricing$ (\emph{left}), $\OKTO$ with $\pricing$
(\emph{middle}) and $\OKTO$
with \emph{DRP} (\emph{right}) as a function of the oversubscription.}
      \label{fig:oversub}
  \end{figure}

Figure~\ref{fig:oversub} shows the impact of the oversubscription factor on the embedded number of virtual
machines. For all
oversubscription ratios, we can observe that $\ALG$ with $\pricing$ outperforms the other combinations, while $\pricing$
is
superior to \emph{DRP} in combination with $\OKTO$. While the differences are small for an oversubscription factor of
$1$, we see an increase of the difference  with the oversubscription factor; it diminishes after an oversubscription
factor of $5$, where the results become stable.

\begin{figure}[ht]
  \centering
\includegraphics[width=.95\columnwidth]{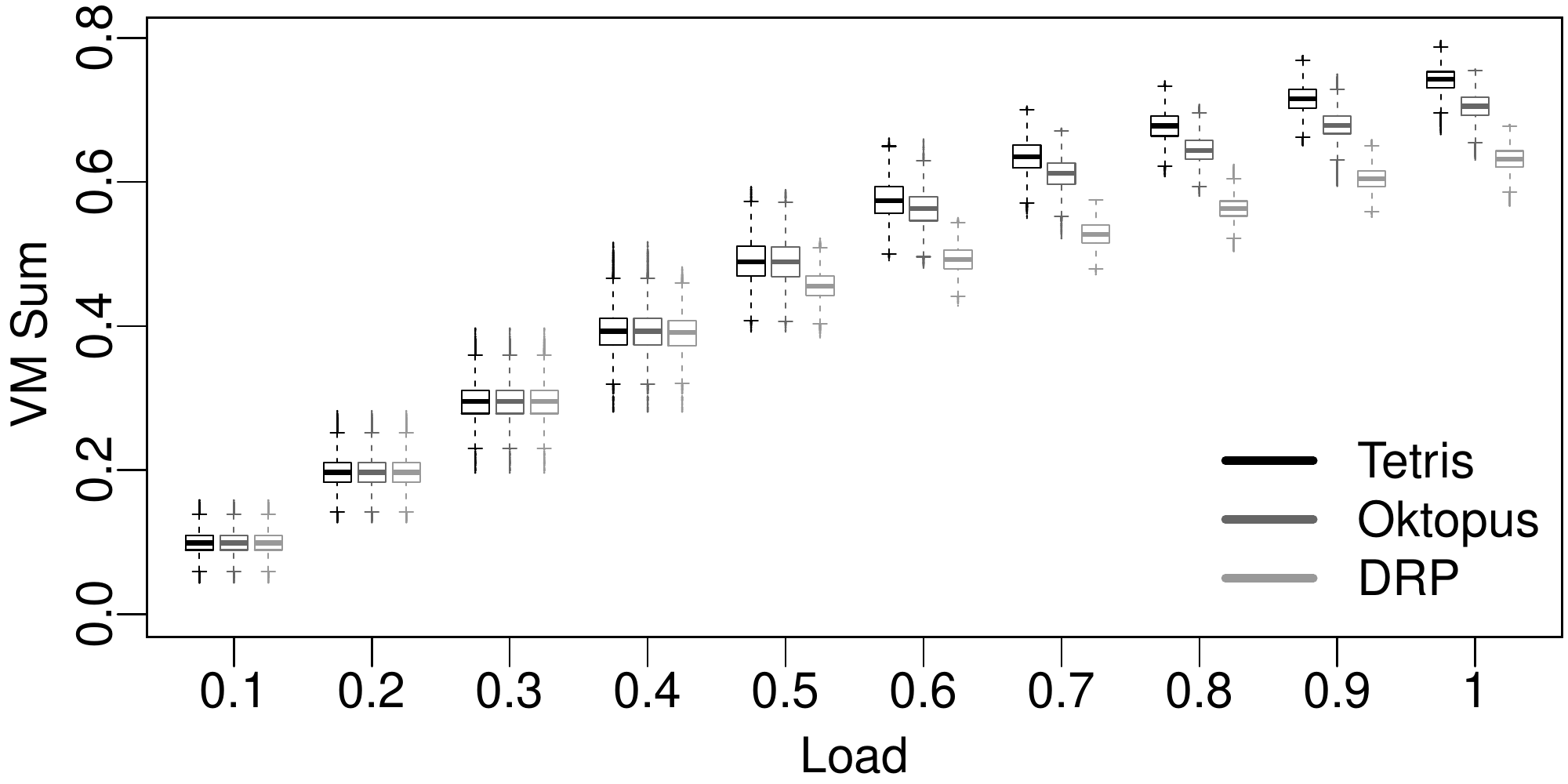}
      \caption{Embedded \emph{VM} slots for $\ALG$ with $\pricing$ (\emph{left}), $\OKTO$ with $\pricing$ (\emph{middle}) and $\OKTO$
with \emph{DRP} (\emph{right}) as a function of datacenter load.}
      \label{fig:load}
  \end{figure}

Naturally, we can only reap the full benefits of $\pricing$ and $\ALG$ in highly utilized datacenters,
as shown in Figure~\ref{fig:load}: While first marginal effects can be observed at a load of $0.4$, the benefits of $\pricing$ are
visible starting at $0.5$ and the benefits of $\ALG$ at $0.6$. The effects increase until a load of $1$.
Given that highly utilized datacenters are a reality today and the key to provider benefits, these results are encouraging.

To analyze the benefit of the pricing model, we consider our default scenario:
The mean resource sum for $\pricing$ with $\OKTO$ is $15\%$ higher than for \emph{DRP} with $\OKTO$. This means that the
amount of concurrent active guarantees is $15\%$ larger. Using $\ALG$ with $\pricing$ yields another $5\%$ improvement,
resulting in a total benefit of $20\%$. Similar numbers apply for the bandwidth resource sum.

Assuming $p_b=p_c$ and a uniform distribution of accepted requests (i.e., an equal amount of embedded \emph{VM}s
of each $(c,b)$ tuple), inserting the $20\%$ as a $\Delta$ in
Section~\ref{sec:pricing} leads to savings of $27\%$ for customers with skewed requests. The corresponding values of
$\lambda_c$ and $\lambda_b$ are $1/6$, i.e., customers have to be charged for about $17\%$ of the resource
difference to \emph{DRP}s templates in order to keep the revenue for the provider constant.

\section{Conclusion}\label{sec:conclusion}

This paper has presented the first specification-dependent pricing scheme for virtual clusters,
the standard abstraction of cloud applications today. Together with the pricing scheme,
we also developed a new virtual cluster embedding algorithm, which may be of independent interest.
Our approach can easily be combined with interfaces which translate high-level
customer goals into virtual cluster requirements and compute
resource combinations, such as~\cite{short-gapbridge}.

We find that the proposed specification-dependent pricing can increase the social welfare,
leading to savings of up to
$25\%$ compared to \emph{DRP}. In fact $\pricing$ enables customer to have guaranteed application performance while they
only need to pay for the resources they use, including a small extra fee.
Moreover, we find that $\ALG$ distributes virtual clusters with skewed resource requirements over several
hosts and therefore embeds $5\%$ more virtual resources than $\OKTO$.

In our future research, we want to extend the $\pricing$ scheme
by a spot market to deal with more volatile traffic patterns.


\end{document}